\documentclass[twocolumn,showpacs,preprintnumbers,amsmath,amssymb]{revtex4}
\usepackage{graphicx}
\usepackage{dcolumn}
\usepackage{bm}

\begin{document}

\title{Plykin-like attractor in non-autonomous coupled oscillators}

\author{Sergey P. Kuznetsov }
\altaffiliation[
Kotel'nikov's Institute of Radio-Engineering and Electronics of RAS, Saratov 
Branch, Zelenaya 38, Saratov, 
410019, Russian Federation.]{}

\date{\today}

\begin{abstract}
A system of two coupled non-autonomous oscillators is considered. Dynamics 
of complex amplitudes is governed by differential equations with periodic 
piecewise continuous dependence of the coefficients on time. The 
Poincar\'{e} map is derived explicitly. With exclusion of the overall phase, 
on which the evolution of other variables does not depend, the Poincar\'{e} 
map is reduced to 3D mapping. It possesses an attractor of Plykin type 
located on an invariant sphere. Computer verification of the cone criterion 
confirms the hyperbolic nature of the attractor in the 3D map. Some results 
of numerical studies of the dynamics for the coupled oscillators are 
presented, including the attractor portraits, Lyapunov exponents, and the 
power spectral density. 
\end{abstract}

\maketitle

\textbf{
In mathematical theory of dynamical systems a class of uniformly hyperbolic 
strange attractors is known. In such an attractor all orbits are of the 
same saddle type, they manifest strong stochastic properties and allow detailed 
theoretical analysis. The mathematical theory was advanced more than 
40 years ago, but till now the hyperbolic strange attractors 
are regarded rather as purified image of deterministic chaos than as realistic models 
of complex dynamics. In textbooks and reviews, examples of 
these attractors are traditionally represented by 
abstract artificial constructions like
the Plykin attractor and the Smale - Williams attractor. 
Recently, a realistic system was suggested and implemented as electronic device, 
dynamics of which in stroboscopic description is associated with attractor 
of Smale - Williams type. 
In the present article I show how the dynamics related to an attractor of 
Plykin type may be obtained in coupled non-autonomous 
self-oscillators. As systems of coupled oscillators occur in many fields 
in physics and technology, it is natural to expect that the suggested model 
may be realizable, for example, with electronic devices, mechanical systems, 
objects of laser physics and nonlinear optics. The systems with hyperbolic 
strange attractors may be of special interest in applications (e.g. for 
noise generators, chaos communication etc.) due to the intrinsic structural 
stability, that means insensitivity of the chaotic motions to variations of 
parameters, characteristics of elements, technical fluctuations etc. 
}

\section{Introduction}
\label{sec1}

Mathematical theory of dynamical systems introduces a class of \textit{uniformly 
hyperbolic strange attractors} \cite{1,2,3,4,5,6,7,8,9}. In 
such an attractor all orbits are of saddle type, and their stable and 
unstable manifolds do not touch, but can only intersect 
transversally. These attractors manifest strong stochastic properties and 
allow a detailed mathematical analysis. They are structurally stable; that 
means insensitivity of the structure of the attractors in respect to 
variation of functions and parameters in the dynamical equations. 
Until very recent times, the hyperbolic strange attractors 
were regarded rather as purified image of chaos than as objects relating to
complex dynamics of real-world systems. (See discussion of the question in 
Ref. \cite{9}; also, a mechanical system with hyperbolic dynamics, 
the so-called triple linkage, has been considered there.)

In textbooks and reviews, examples of the uniformly hyperbolic attractors 
are traditionally represented by mathematical constructions, the Plykin 
attractor and the Smale -- Williams solenoid. These examples relate to 
discrete-time systems, the iterated maps. The Smale -- Williams attractor 
appears in the mapping of a toroidal domain into itself in the state space 
of dimension 3 or more. The Plykin attractor occurs in some special mapping 
on a sphere with four holes, or in a bounded domain on a plane with three 
holes (Fig.~1~a) \cite{10}. It is known that a variety of topologically different 
Plykin-like attractors may be constructed in finite two-dimensional domains 
with holes. One of the modifications shown in Fig.~1~b is of special interest 
for the present study and will be referred to as the Plykin -- Newhouse 
attractor \cite{11,4}. 

In applied disciplines, physics and technology, people deal more often with 
systems operating in continuous time; they are called the flows in 
mathematical literature. The procedure of passage from mapping 
${\rm {\bf x}}_{n + 1} = {\rm {\bf f}}({\rm {\bf x}}_n )$ 
to a flow system is called 
\textit{suspension} \cite{2,3,4,5,6,7}. 
Such a passage is possible if the map is reversible. For the 
resulting flow system the relation 
${\rm {\bf x}}_{n + 1} = {\rm {\bf f}}({\rm {\bf x}}_n )$ 
is the Poincar\'{e} map, which in the context on non-autonomous systems is called
sometimes the stroboscopic map.

Recently, a system was suggested and realized experimentally, in which the 
Poincar\'{e} map possesses an attractor of Smale -- Williams type \cite{12, 13}. 
It is composed of two non-autonomous van der Pol oscillators, which become 
active turn by turn and transfer the excitation each other, in such manner that the 
transformation of the phase of oscillations on a whole cycle corresponds to 
expanding circle map. Computer verification of conditions guaranteeing 
the hyperbolic nature of the attractor was performed in Ref. \cite{14}. 
(See some developments of the scheme in Refs.~\cite{14a,14b,14c,14d}.) 

Till now, no explicit examples were advanced for a Plykin type attractor to 
occur in a low-dimensional physically realizable system \footnote{ A special 
comment is needed to the work of Halbert and Yorke \cite{15} announcing a 
physical realization of the Plykin attractor. As a physical object, the 
taffy-pulling machine they discuss is not a low-dimensional system, but 
contains a piece of continuous medium undergoing deformations in such way 
that the motion of local elements of the medium obeys a map with the 
Plykin-like attractor. In other words, it is an ensemble of elements, each 
of which carries out motion on the Plykin-like attractor. Thus, referring to 
the physical realization of the attractor, the authors stand for another 
meaning than that we have in mind here (as well as other authors 
\cite{9,16,17,18}). }. In Refs. \cite{16} and \cite{17} the authors argue in favor of 
existence of the Plykin-type attractors in the Poincar\'{e} maps for a 
modified Lorenz system and for an autonomous three-dimensional system 
modeling dynamics of neuron. On the other hand, an explicit example of a 
non-autonomous flow system with Plykin-Newhouse attractor in the 
stroboscopic map has been advanced in the PhD thesis of Hunt \cite{18}. The 
model of Hunt is defined by multiple expressions, distinct for different 
domains in the state space, and contains many artificially introduced 
smoothing factors. It is really hard to imagine that this model could be 
reproduced on a base of some physical system. 
\begin{figure}[htbp]
\includegraphics[width=3in]{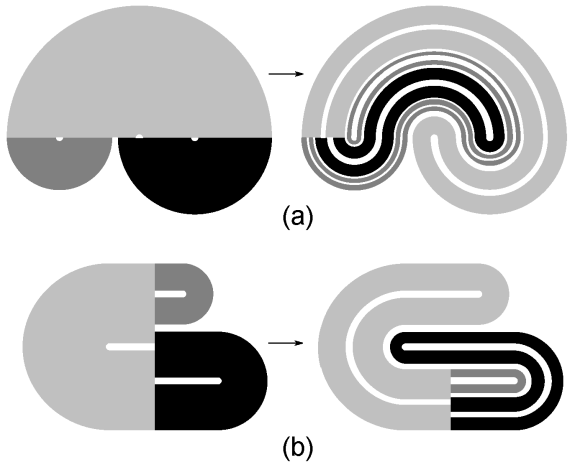}
\label{figur1}
\caption{
Illustration of action on a plane for the map suggested in the 
original paper of Plykin (a), and for the version of the map with the Plykin 
-- Newhouse attractor (b). Each of them may be associated with a map defined 
on the sphere, say, by means of the stereographic projection
}
\end{figure}

In the present article I show how the dynamics associated with attractor of 
Plykin type may be obtained in a system of coupled non-autonomous 
oscillators. As believed, it opens prospects for constructing physical and 
technical systems, e.g. electron devices with the structurally stable 
chaotic regimes. 

In Section \ref{sec2} a sequence of continuous transformations is defined on a 
two-dimensional sphere, and a system of two coupled oscillators is 
introduced, in which the state evolution corresponds in some sense to those 
transformations. The equations are written down for complex amplitudes of 
the oscillations. The points on the sphere represent the instantaneous 
states defined up to the overall phase factor. An explicit Poincar\'{e} map 
is derived that describes evolution of the state on one period of variation 
of coefficients in the non-autonomous differential equations. With exclusion 
of the overall phase, on which the evolution of other variables does not 
depend, the Poincar\'{e} map is reduced to a three-dimensional map, which 
possesses an attractor of Plykin type on an invariant sphere. In Section \ref{sec3} 
results of computer verification of the so-called cone criterion are 
presented confirming the hyperbolic nature of the attractor of the 
three-dimensional map; it means also its structural stability. The 
topological type of the attractor corresponds to the construction of Plykin 
-- Newhouse. In Section \ref{sec4} some results of numerical studies of the dynamics 
of the coupled oscillator system are discussed, including portraits of the 
attractor, Lyapunov exponents, power spectral density. In the set of 
equations for complex amplitudes, because of presence of a neutral direction 
in the state space, which is associated with the overall phase, the 
attractor has to be related formally to the class of partially hyperbolic 
ones \cite{19,6}. 

\section{Representation of states on a~sphere and equations 
describing dynamics of the model}
\label{sec2}

Let us start with a system of two self-oscillators with compensation of 
losses from the common energy source. Let the equations for the slow 
amplitudes $a$ and $b$ read
\begin{equation}
\label{eq1}
\begin{array}{l}
 \dot {a} = \textstyle{1 \over 2}\mu (1 - \vert a\vert ^2 - \vert b\vert 
^2)a, \\ \\ 
 \dot {b} = \textstyle{1 \over 2}\mu (1 - \vert a\vert ^2 - \vert b\vert 
^2)b, \\ 
 \end{array}
\end{equation}
where $\mu $ is a positive parameter. Let us set
\begin{equation}
\label{eq1a}
\begin{array}{l}
b = \sqrt \rho e^{i\varphi / 2 + i\psi }\sin (\theta / 2),
\\ \\ 
a = \sqrt \rho e^{ - i\varphi / 2 + i\psi }\cos (\theta / 2).
\\ 
\end{array}
\end{equation}
Clearly, in sustained regime of self-oscillations, the condition 
$\rho = \vert a\vert ^2 + \vert b\vert ^2 = 1$ 
has to be valid. If we consider states satisfying $\rho =1$  and identify 
the states differing only in the overall phase, 
we can associate them with the points on a unit sphere 
(Fig.~2). Also, on the picture the Cartesian coordinates are shown:
\begin{equation}
\label{eq2}
\begin{array}{l}
x = \rho \cos \varphi \sin \theta,
\\
y = \rho \sin \varphi \sin \theta,  
\\
z = \rho \cos \theta.
\\ 
\end{array}
\end{equation}
Via the complex amplitudes they are expressed as 
\begin{equation}
\label{eq3}
x + iy = 2a^\ast b,\,\,\,z = \vert a\vert ^2 - \vert b\vert ^2.
\end{equation}

We intend to modify the model (\ref{eq1}) in order to obtain a set of equations with 
coefficients periodically varying in time, in such way that in the 
stroboscopic description and in the representation on the unit sphere, the 
Plykin type attractor will occur. 

As proved by Plykin, a uniformly hyperbolic attractor may exist on a sphere 
only in presence of at least four holes, the areas not visiting by 
trajectories belonging to the attractor. In our construction, the holes will 
correspond to neighborhoods of four points A, B, C, D, with 
coordinates $(x,y,z) = (\pm 1 / \sqrt 2 ,\,\,0,\,\,\pm 1 / \sqrt 2 )$. 

Let us consider a sequence of the following continuous transformations on 
the sphere: 
\begin{itemize}
\item
\textbf{Flow down along circles of latitude,} that is motion of the 
representative points on the sphere away from the meridians NABS and NDCS 
towards the meridians equally distant from the arcs AB and CD.
\item
\textbf{Differential rotation} around $z$-axis with angular velocity depending 
on z linearly, in such way that the points B and C do not move, while the 
points A and D exchange their location.
\item
\textbf{Rotation} of the sphere by $90^{\circ}$
around the $y$-axis.
\item
\textbf{Flow down along circles of latitude}, like at the first stage. 
\item
\textbf{Inverse differential rotation }around $z$-axis.
\item
\textbf{Inverse rotation }by $90^{\circ}$ around the $y$-axis.
\end{itemize}

The procedure is symmetric in the sense that the operations for the stages 
(I) and (IV) are identical, while the stages (II) and (III) differ from (V) 
and (VI) only by directions of the rotations. Intuitively, it looks 
reasonable that this sequence of transformations will generate a flow on the 
sphere accompanying with formation of filaments of fine transversal 
structure, presence of which is a characteristic feature of the Plykin type 
attractors.
\begin{figure}[htbp]
\includegraphics[width=2.5in]{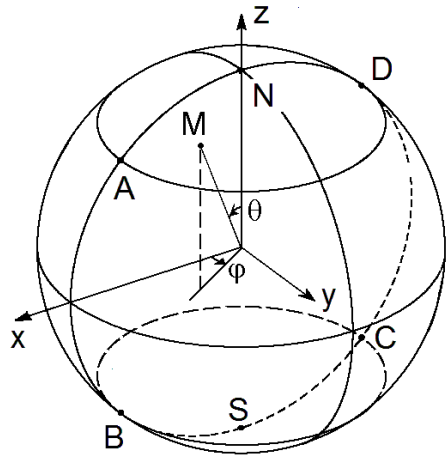}
\label{figur2}
\caption{
The unit sphere with marked points A, B, C, D, neighborhoods 
of which in the further construction will correspond to the holes not 
visiting by trajectories on the attractor. The north and south poles are 
indicated with N and S, respectively. The angular coordinates 
($\theta $, $\varphi )$ 
are shown for some point M, and axes of the Cartesian 
coordinates $x$, $y$, $z$ are depicted
}
\end{figure}

Let us construct equations for the complex amplitudes  
to reproduce dynamics on the above stages with corresponding 
motion of the points on the unit square representing the instantaneous 
states of the system. Duration of each of the six stages is accepted to be 
equal to a unit time interval.

On a stage of \textbf{flow down along circles of latitude} we require the 
angular velocity of motion on the sphere to be proportional to
$\sin 2\varphi$. 
The simplest appropriate form of the differential equations is
\begin{equation}
\label{eq4}
\dot {a} = - i\varepsilon a \, \mbox{Im}({a^*}^2 b^2),\,\,\,
\dot {b} = i\varepsilon b \,\mbox{Im}({a^*}^2 b^2).
\end{equation}
Indeed, substituting $b = e^{i\varphi / 2 + i\psi }\sin (\theta / 2)$ and 
$a = e^{ - i\varphi / 2 + i\psi }\cos (\theta / 2)$, after some simple 
transformations we get 
$\dot {\varphi } = \textstyle{1 \over 2}\varepsilon 
\sin ^2\theta \sin 2\varphi ,\,\,\dot {\theta } = 0$. 
Physically, the terms 
in the right-hand parts of (\ref{eq4}) give rise to a frequency shift 
of opposite 
sign for two oscillators. Magnitude of the shift is proportional to the 
amplitude of a low-frequency signal generated by mixing of the second 
harmonic components from the oscillators on a quadrtatic nonlinear element. 

On a stage of \textbf{differential rotation,} we set 
\begin{equation}
\label{eq5}
\begin{array}{l}
\dot {a} = \textstyle{1 \over 4}i\sigma \pi (\sqrt 2 - 1 - 2\sqrt 2 \vert 
a\vert ^2)a,
\\ \\
\dot {b} = \textstyle{1 \over 4}i\sigma \pi (\sqrt 2 + 1 
- 2\sqrt 2 \vert b\vert ^2)b,
\end{array}
\end{equation}
where $\sigma = \pm 1$. In angular variables $(\varphi ,\,\,\theta )$, these 
equations reduce to 
$\dot {\varphi } = \textstyle{1 \over 2}\sigma \pi 
(\sqrt 2 \cos \theta + 1)$, $\,\dot {\theta } = 0$. 
Note that the angular 
velocity $\dot {\varphi }$ depends linearly on $z = \cos \theta $ and 
vanishes at $z = - 1 / \sqrt 2 $. On this stage two subsystems must behave 
like uncoupled classic non-isochronous oscillators. At small amplitudes 
their frequencies in respect to the reference point are 
$\Delta \omega _{a,b} = \textstyle{1 \over 4}i\sigma \pi (\sqrt 2 \mp 1)$. 
With growth of 
the amplitudes, the oscillation frequencies undergo a shift proportional to 
the squared amplitude for both subsystems.

Finally, on the stages of \textbf{rotation} an appropriate form of the 
equations is
\begin{equation}
\label{eq6}
\dot {a} = - \textstyle{1 \over 4}\pi sb,\,\,\,\dot {b} = \textstyle{1 \over 
4}\pi sa,
\end{equation}
where $s = \pm 1$. It corresponds to a conservative system of coupled 
oscillators with equal frequencies, with the coupling coefficient of such 
value that the energy exchange between the partial oscillators corresponds 
precisely to the duration of the stage. 

Now, we can write down equations for the complex amplitudes embracing the 
complete time period $T=6$. For this, we compose the right-hand sides as 
combinations of terms (\ref{eq4}), (\ref{eq5}), (\ref{eq6}), which are supposed to be switched in, 
or off, during the respective stages of the time evolution. As to the terms 
from the equations (\ref{eq1}), we account them only on the stages of rotation. 
(Their exclusion for other stages is not so significant, but we do so, 
because it simplifies derivation of the Poincar\'{e} map in the explicit 
form.) Finally, we arrive at the equations 
\begin{equation}
\label{eq7}
\begin{array}{c}
\dot {a} = - i\varepsilon (1 - \sigma ^2 - s^2)\,\mbox{Im}(a^2b^{\ast 2})a-\textstyle{\pi \over 4}sb  \\ 
+\textstyle{1 \over 4}i\sigma \pi (\sqrt 2 - 1 - 2\sqrt 2 \vert a\vert ^2)a 
\\
+ \textstyle{1 \over 2}s^2\mu (1 - \vert a\vert ^2 
- \vert b\vert ^2)a, 
\\ \\
\dot {b} = i\varepsilon (1 - \sigma ^2 - s^2)\,\mbox{Im}(a^2b^{\ast 2})b + \textstyle{\pi \over 4}sa \\
+ \textstyle{1 \over 4}i\sigma \pi (\sqrt 2 + 1 - 2\sqrt 2 \vert b\vert ^2)b 
\\ 
+ \textstyle{1 \over 2}s^2\mu (1 - \vert a\vert ^2 
- \vert b\vert ^2)b. \\ 
 \end{array}
\end{equation}
Here the factors $\sigma $ and $s$ depend on time with period $T = 6$, and on a 
single period they are defined by the relations 
\begin{equation}
\label{eq8}
\sigma = \left\{ {{\begin{array}{*{20}c}
 { - 1,\,\,1 \le t < 2} \hfill \\
 {\,\,\,\,\,\,1,\,\,4 \le t < 5,} \hfill \\
 {\,\,\,\,\,\,0,\,\,{\rm otherwise,}} \hfill \\
\end{array} }} \right.\,\,\,s = \left\{ 
{{\begin{array}{*{20}c}
 { - 1,\,\,2 \le t < 3} \hfill \\
 {\,\,\,\,\,\,1,\,\,5 \le t < 6,} \hfill \\
 {\,\,\,\,\,\,0,\,\,{\rm otherwise.}} \hfill \\
\end{array} }} \right.
\end{equation}

Let us derive the Poincar\'{e} map, which determines transformation of the 
state over one period $T=6$ and describes the time evolution stroboscopically. 

Let the initial conditions for the equations (\ref{eq2}) at $t_n = nT$ are defined 
as the state vector ${\rm {\bf X}}_n = (a_n ,b_n )$, and the state after a 
half of period is 
${\rm {\bf X}}_{n + 1 / 2} = {\rm {\bf F}}_{\sigma ,s} 
({\rm {\bf X}}_n ) \quad  = (a_{n + 1 / 2} ,b_{n + 1 / 2} )$.

Solving equations (\ref{eq7}) on each successive unit interval 
analytically, we can represent the map ${\rm {\bf F}}_{\sigma ,s} $ 
explicitly:
\begin{equation}
\label{eq9}
\begin{array}{l}
 a_{n + 1/2} = \frac{a_n 
De^{i\alpha + \mu / 2} - sb_n D^\ast e^{i\beta + \mu / 2}}{\sqrt {1 + (\vert 
a_n \vert ^2 + \vert b_n \vert ^2)(e^\mu - 1)} }\,\,, \\ \\
 b_{n + 1/2} = \frac{sa_n 
De^{i\alpha + \mu / 2} + b_n D^\ast e^{i\beta + \mu / 2}}{\sqrt {1 + (\vert 
a_n \vert ^2 + \vert b_n \vert ^2)(e^\mu - 1)} }\,\,, \\ 
 \end{array}
\end{equation}
where
\[
\begin{array}{c}
 \alpha = \textstyle{1 \over 4}\sigma \pi (\sqrt 2 - 1 - 2\sqrt 2 \vert a_n 
\vert ^2),\,\,\, \\ \\
 \beta = \textstyle{1 \over 4}\sigma \pi (\sqrt 2 + 1 - 2\sqrt 2 \vert b_n 
\vert ^2), \\ \\ 
D = \frac{1}{\sqrt 2 }\,\left( {\frac{\vert a_n \vert ^2\vert b_n \vert ^2 
- (a_n^ * b_n )^2\tanh (2\varepsilon \vert a_n \vert ^2\vert b_n \vert 
^2)}{\vert a_n \vert ^2\vert b_n \vert ^2 - (a_n b_n^ * )^2\tanh 
(2\varepsilon \vert a_n \vert ^2\vert b_n \vert ^2)}} 
\right)^{1/4}. \\ 
\end{array}
\]
The indices $\sigma $ and $s$ become equal $\pm 1$ alternately, so the mapping 
for the complete period is defined as follows: 
\begin{equation}
\label{eq11}
{\rm {\bf X}}_{n + 1} = {\rm {\bf F}}({\rm {\bf X}}_n ) = {\rm {\bf 
F}}_{1,1} ({\rm {\bf F}}_{ - 1, - 1} ({\rm {\bf X}}_n )).
\end{equation}

Dynamics governed by equations~(\ref{eq7}) or by iterations of the map (\ref{eq11}) is 
invariant in respect to simultaneous constant phase shift for two 
oscillators, i.e. to the variable change 
$a \to ae^{i\psi },\,\,b \to be^{i\psi }$. 
Due to this, one can reduce the equations for two complex 
amplitudes to equations in three real variables. 

Performing the variable change (\ref{eq3}), we get 
a set of differential equations
\begin{equation}
\label{eq12}
\begin{array}{l}
\begin{array}{r}
\dot {x} = - \textstyle{1 \over 2}\,\sigma \pi (z\sqrt 2 + 1)y 
- \varepsilon (1 - \sigma ^2 - s^2)xy^2 
\\ + \textstyle{1 \over 2}s\pi z 
+\mu s^2(1 - \sqrt {x^2 + y^2 + z^2} )x, 
\end{array}
\\ \\
\begin{array}{r}
\dot {y} = \textstyle{1 \over 2}\sigma \pi (z\sqrt 2 + 1)x / 2 
+ \varepsilon (1 - \sigma ^2 - s^2)yx^2 \\
+\mu s^2(1 - \sqrt {x^2 + y^2 + z^2} )y, 
\end{array}
\\ \\
\dot {z} =  
- \textstyle{1 \over 2}s\pi x + \mu s^2(1 - \sqrt {x^2 + y^2 + z^2} )z, \\ 
\end{array}
\end{equation}
where $\sigma $ and $s$ are time-dependent, as stated by formulas (\ref{eq8}). 
Designating at $t_n = nT$ the state vector as 
${\rm {\bf x}}_n = (x_n ,y_n ,z_n )$, 
we can represent the three-dimensional Poincar\'{e} map as
\begin{equation}
\label{eq13}
{\rm {\bf x}}_{n + 1} = {\rm {\bf f}}({\rm {\bf x}}_n ) = {\rm {\bf 
f}}_{1,1} ({\rm {\bf f}}_{ - 1, - 1} ({\rm {\bf x}}_n )),
\end{equation}
and the half-period map 
${\rm {\bf x}}_{n + 1 / 2} = {\rm {\bf f}}_{\sigma ,s} ({\rm {\bf x}}_n )$ 
is expressed as
\begin{equation}
\label{eq14}
\begin{array}{c}
 x_{n + 1 / 2} = Psz_n , \\ \\ 
 y_{n + 1 / 2} = PQ\left[ {\sigma x_n e^{ - \varepsilon (x_n^2 + y_n^2 ) / 
2}\sin \textstyle{\pi \over 2}(z_n \sqrt 2 + 1)} \right. \\ 
 \left. { + y_n e^{\varepsilon (x_n^2 + y_n^2 ) / 2}\cos \textstyle{\pi 
\over 2}(z_n \sqrt 2 + 1)} \right], \\ \\ 
 \,z_{n + 1 / 2} = PQ\left[ { - sx_n e^{ - \varepsilon (x_n^2 + y_n^2 ) / 
2}\cos \textstyle{\pi \over 2}(z_n \sqrt 2 + 1)} \right. \\ 
 \left. { + s\sigma y_n e^{\varepsilon (x_n^2 + y_n^2 ) / 2}\sin 
\textstyle{\pi \over 2}(z_n \sqrt 2 + 1)} \right], \\ 
 \end{array}
\end{equation}
where
\[
\begin{array}{c}
 P = \left[ {(1 - e^{ - \mu })\sqrt {x_n^2 + y_n^2 + z_n^2 } + e^{ - \mu}} \right]^{ - 1}, \\ \\ 
 Q = \frac{\sqrt {x_n^2 + y_n^2 } }{\sqrt {x_n^2 e^{ - \varepsilon (x_n^2 + 
y_n^2 )} + y_n^2 e^{\varepsilon (x_n^2 + y_n^2 )}} }. \\ 
 \end{array}
\]

The maps (\ref{eq11}) and (\ref{eq13}) are reversible. The inverse maps are derived from 
solution of the equations (\ref{eq7}) or (\ref{eq12}) in the backward time; their analytic 
representations are omitted for brevity. 

\section{Numerical results for the~three-dimensional map 
and hyperbolic nature of the attractor }
\label{sec3}

In Fig.~3~a-c portraits of the attractor are shown for the map 
${\rm {\bf x}}_{n + 1} = {\rm {\bf f}}({\rm {\bf x}}_n )$ 
at $\varepsilon =1$, $\mu =1$. 
They are obtained by computation of a sufficiently large number of 
iterations after excluding the initial transient part of the orbit. As seen 
from the diagram (a), in the space ($x$,$y$,$z$) the attractor is disposed on a unit 
sphere. In the diagram (b) it is represented in the angular 
coordinates $(\varphi ,\,\,\theta )$, and in the diagram (c) as an object on 
the plane of the complex variable

\begin{figure*}[htbp]
\includegraphics[width=6in]{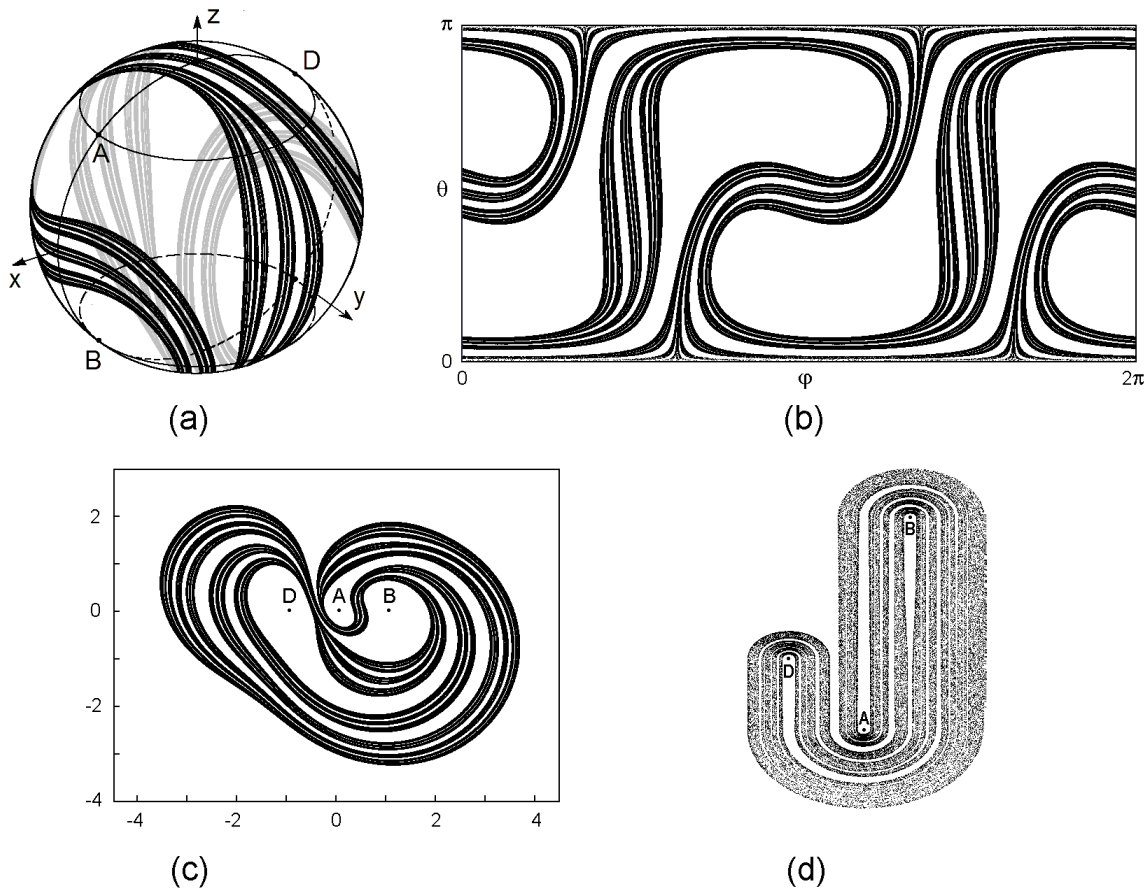}
\label{figur3}
\caption{
Attractor of the map (\ref{eq13}) at $\varepsilon =1$ and $\mu =1$ 
in three-dimensional space on the unit square (a), its representation in the 
angular coordinates $(\varphi ,\,\,\theta )$ (b), and a portrait on the plane 
in stereographic projection (c). Panel (d) shows for comparison the portrait 
of the Plykin -- Newhouse attractor reproduced from Ref. \cite{20} for the Hunt 
model \cite{18}. The orientation is selected specially to see better the 
correspondence of the structure of the filaments with that in the diagram 
(c) 
}
\end{figure*}

\begin{equation}
\label{eq15}
W = \frac{x - z + iy\sqrt 2 }{x + z + \sqrt 2 }.
\end{equation}

The last corresponds to stereographic projection from the sphere to the 
plane with a use of the point $C$ as a center. This point together with a 
neighborhood does not belong to the attractor of the map; so, its image occupies a 
bounded part of the plane $W$. 

Note specific fractal-like transverse structure of the attractor. Few 
initial levels of this structure are easily distinguishable: the object 
looks like composed of strips, each of which contains narrower strips of the 
next level etc. As follows from the computations discussed below, it is \textit{a uniformly hyperbolic strange attractor}. 
Its topological type corresponds to the Plykin -- Newhouse attractor. The 
last follows from visual comparison of mutual location of filaments in the 
diagram (c) and in the Plykin -- Newhouse attractor shown in diagram (d). 
(The last picture is taken from the paper \cite{20}, which reproduces analysis of 
the Hunt model \cite{18}, definitely possessing the attractor of Plykin -- 
Newhouse.) 

To compute all Lyapunov exponents for the three-dimensional map, joint 
iterations of (\ref{eq13}) together with a collection of three equations in 
variations for perturbation vectors are produced. At each step, 
Gram--Schmidt process is applied to obtain an orthogonal set of vectors, and 
normalization of the vectors to a fixed constant is performed. Lyapunov 
exponents are obtained as slopes of the straight lines approximating the 
accumulating sums of logarithms of the norm ratios for the vectors in 
dependence of the number of iterations \cite{21}. In particular, at 
$\varepsilon =1$ and $\mu =1$ the Lyapunov exponents are, 
$\Lambda _1 = 0.9575$, 
$\Lambda _2 = - 1.2520$, 
$\Lambda _3 = - 2$, 
and estimate of the attractor dimension 
with the Kaplan -- Yorke formula yields 
$D_L = 1 + \Lambda _1 / \vert \Lambda _2 \vert \approx 1.765$.

To substantiate the hyperbolic nature of the attractor, let us turn to 
computational verification 
of the \textit{cone criterion} known from the mathematical literature 
\cite{6,7,8,22,18,14}.

Let us have a smooth map 
${\rm {\bf \bar {x}}} = {\rm {\bf g}}({\rm {\bf x}})$ 
that determines discrete-time dynamics on an attractor $A$.

The criterion requires that with appropriate selection of a constant 
$\gamma > 1$, for any point ${\rm {\bf x}} \in A$, in the space of vectors of 
infinitesimal perturbations one can define the expanding and contracting 
cones $S_{\rm {\bf x}} $ and $C_{\rm {\bf x}} $. Here $S_{\rm {\bf x}} $ 
is a set of vectors satisfying the condition that their norms increase by 
factor $\gamma $ or more under the action of the map. 
$C_{\rm {\bf x}}$ is a set of vectors, 
for which the norms increase by factor 
$\gamma $ or more under the action of the inverse map 
${\rm {\bf \tilde {x}}} = {\rm {\bf g}}^{-1}({\rm {\bf x}})$. 
The cones $S_{\rm {\bf x}}$ and $C_{\rm {\bf x}}$ 
must be invariant in the following sense. (i) For 
any ${\rm {\bf x}} \in A$ the image of the expanding cone from the pre-image 
point ${\rm {\bf \tilde {x}}}$ must be a subset of the expanding cone at 
\textbf{x}. (ii) For any ${\rm {\bf x}} \in A$ the pre-image of the 
contracting cone from the image point ${\rm {\bf \bar {x}}}$ must be a 
subset of the contracting cone at \textbf{x}. 

Let ${\rm {\bf g}}({\rm {\bf x}})$ 
be a map corresponding to the $k$-fold 
iteration of the Poincar\'{e} map 
${\rm {\bf f}}^k({\rm {\bf x}})$, 
where $k$ is an integer selected in the course of the computations. 
The needed 
Jacobian matrices can be found in our case analytically, by differentiating 
of (\ref{eq13}) with application of the chain rule for the derivatives of the 
functional compositions. Some details of the computational procedure, which 
takes into account disposition of the attractor on the invariant sphere, are 
given in Appendix.

The calculations are organized as verification of the required conditions 
for a set of point on the attractor obtained from multiple iterations of the 
map ${\rm {\bf g}}({\rm {\bf x}})$. We check, first, the existence of 
nonempty expanding and contracting cones, and secondly, the validity of 
inequalities corresponding to proper inclusions of these cones. If these 
conditions are met with $\gamma =1$, the interval, is determined 
$\gamma _{\min } ({\rm {\bf x}}) 
\le \gamma \le \gamma _{\max } ({\rm {\bf x}})$, 
in which they are true.

Figure 4 shows a diagram resuming graphically results of verification of the 
cone criterion for the attractor of the map (\ref{eq13}) 
at $\varepsilon =1$ and 
$\mu =1$. The computations have been performed for the map 
${\rm {\bf f}}^3({\rm {\bf x}})$. 
The diagram represents in 
logarithmic scale the values 
$\gamma _{\min } ({\rm {\bf x}})$ 
in gray and 
$\gamma _{\max } ({\rm {\bf x}})$ 
in black versus $y$ coordinate of the analyzed points on the 
attractor. Observe a gray set and a black set, one disposed strongly 
above, and another strongly below the horizontal line $\gamma = 1$. Presence 
of a gap separating these sets from the line $\gamma = 1$ implies the 
positive result of the test. To express the result quantitatively, one can 
determine the maximum of 
$\gamma _{\min } ({\rm {\bf x}})$ 
and minimum of 
$\gamma _{\max } ({\rm {\bf x}})$ 
over the set of all processed 
points. As found, selection of the constant satisfying 
$0.44 < \gamma ^2 < 2.3$ ensures the required invariance of the cones. 

\begin{figure}[htbp]
\includegraphics[width=3.3in]{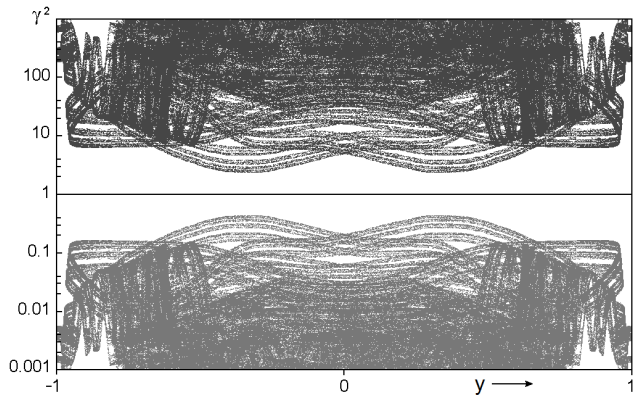}
\label{figur4}
\caption{
A graphical illustration for verification of the 
hyperbolic nature of the attractor for the map (\ref{eq13}), with $\varepsilon $=1 
and $\mu $=1. A positive result of the test follows from existence of the 
gap between the line $\gamma = 1$ and the black and gray sets of points, 
which represent, respectively, the upper and lower edges of the intervals of 
$\gamma $, which met the verified conditions
}
\end{figure}

\section{Numerical results for the coupled oscillators}
\label{sec4}

In accordance with the previous section, there is a correspondence between 
dynamics of complex amplitudes in the coupled oscillators (\ref{eq7}) 
and dynamics 
of the three-dimensional mapping (\ref{eq13}) 
possessing the hyperbolic attractor of 
Plykin -- Newhouse. Let us illustrate with numerical results the dynamics of 
the coupled oscillators concentrating on features linked with the flow 
nature of the system, i.e. with the continuous time evolution. 

\begin{figure*}[htbp]
\includegraphics[width=6in]{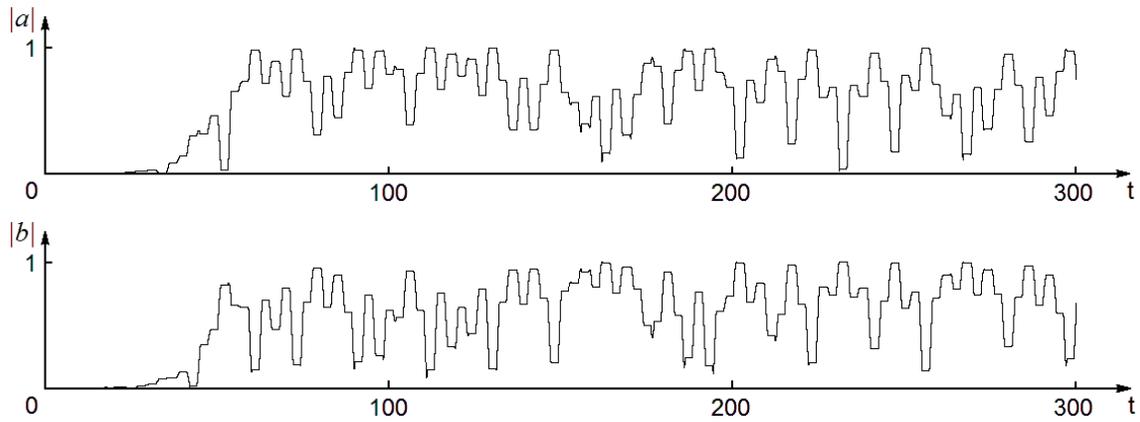}
\label{figur5}
\caption{
Plots of the real amplitudes $\vert a\vert $ and $\vert 
b\vert $ versus time in the transient process obtained from numerical 
solution of the differential equations (\ref{eq7}) 
at $\varepsilon $=1 and \textit{$\mu $}=1
}
\end{figure*}

Figure 5 shows plots for the amplitudes of the coupled oscillators $\vert 
a\vert $ and $\vert b\vert $ versus time obtained from numerical solution of 
the differential equations (\ref{eq7}) with the Runge -- Kutta method at 
$\varepsilon =1$ and $\mu =1$. 
 Some small in absolute value 
and random in phase complex amplitudes $a$ and $b$ are taken as initial conditions, 
so the plot depicts 
the transient process prior to the regime of chaotic self-oscillations. In 
the right-hand part of the diagram the dependences look like samples of a 
random process; that associates with motion on the chaotic attractor. 
Locally, some peculiarities can be seen because of the piecewise continuous 
nature of the process composed of successive stages. In particular, the 
horizontal plateaus relate to the stages of evolution, on which the 
amplitudes $\vert a\vert $ and $\vert b\vert $ remain constant according to 
equations (\ref{eq5}). Note that the realizations 
for $\vert a\vert $ and $\vert 
b\vert $ are interconnected: in the sustained regime they obey the relation 
$\vert a\vert ^2 + \vert b\vert ^2 = 1$.

\begin{figure*}[htbp]
\includegraphics[width=2.5in]{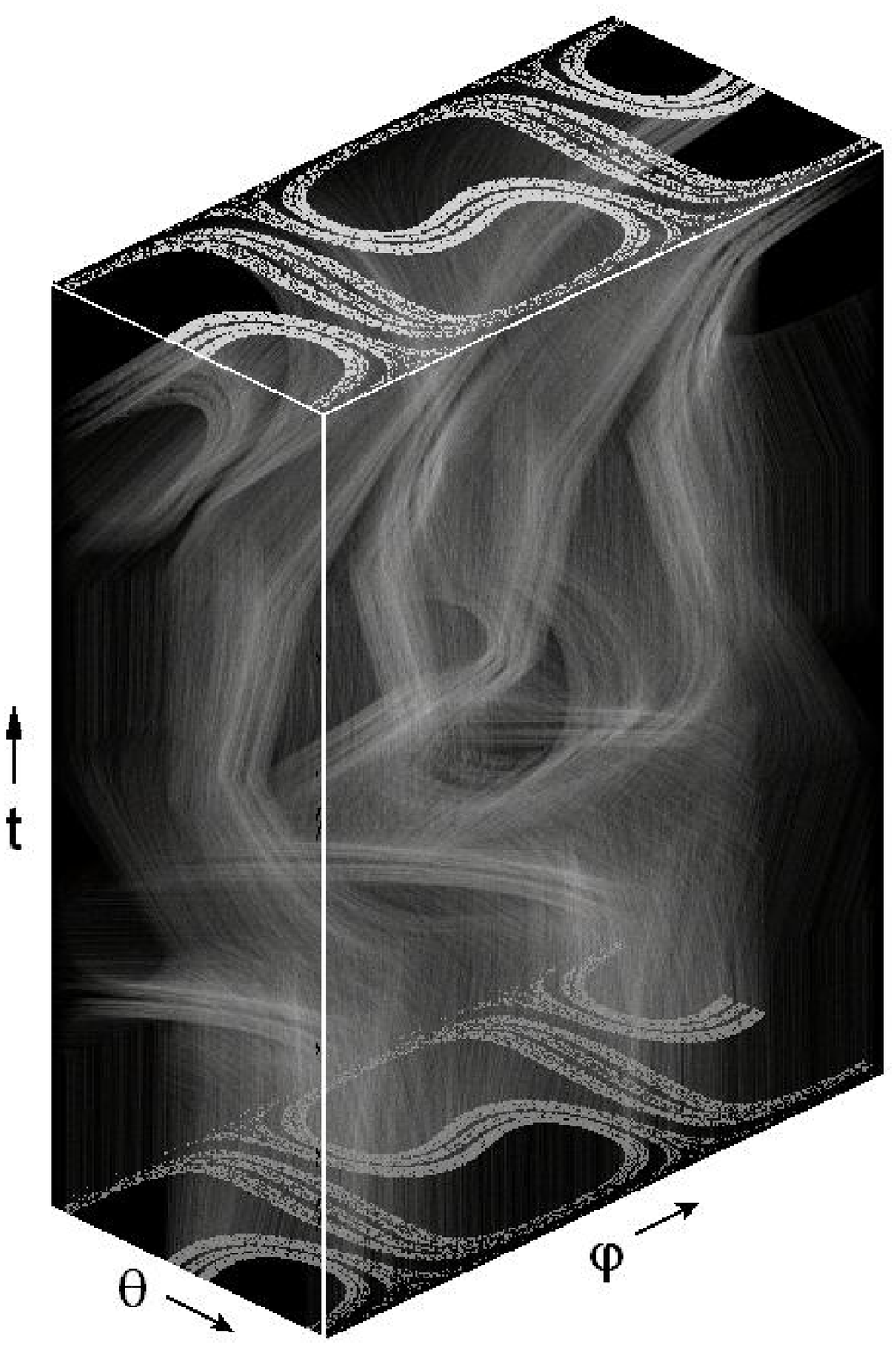}
\includegraphics[width=3.4in]{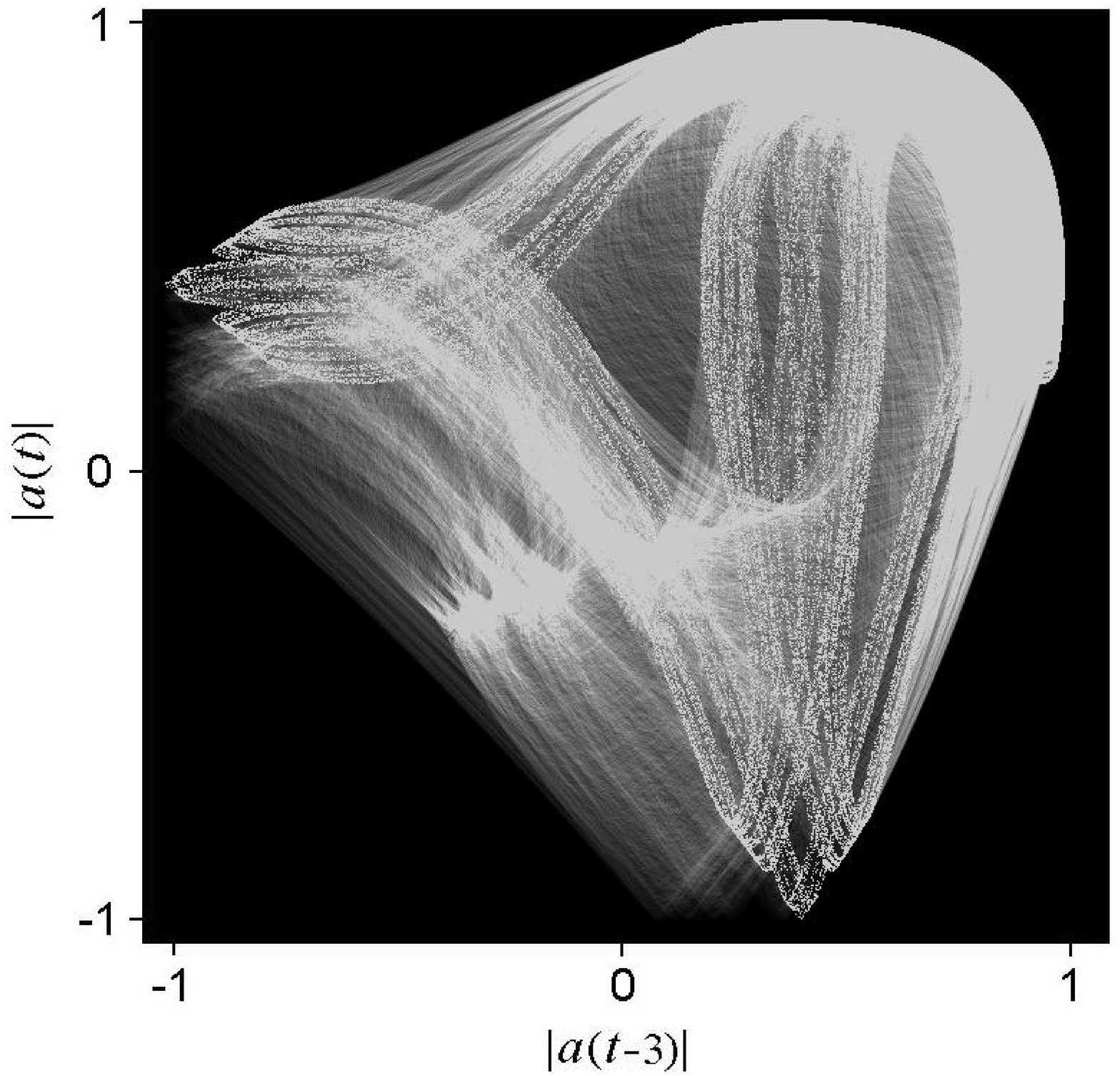}
\label{figur6}
\caption{
(a) Portrait of attractor for the system (\ref{eq7}) in 
three-dimensional space $(\varphi ,\theta ,\,\,t)$. In the cross-section 
with the horizontal plane $t = 0\,\,(\bmod 6)$ observe the object identical 
to that shown in Fig.~3~b. (b) Portrait of the attractor in the plane of real 
amplitudes relating to one of the oscillators and separated by time interval $T$/2=3. 
Technique of representation in gray scales is used: brighter tones 
correspond to higher probability of visiting the pixels by the 
representative points. The parameter values are $\varepsilon $=1, $\mu $=1
}
\end{figure*}

Figure 6 presents two versions of portraits of the attractor for the system 
(\ref{eq7}) at $\varepsilon =1$ and $\mu=1$. 
As the dimension of the state space is 
sufficiently high (vector ${\rm {\bf X}} = (a,b)$ is four-dimensional, and 
the extended state space of the non-autonomous system is five-dimensional), 
depicting the image to resolve subtle fractal transverse structure intrinsic 
to the attractor is not a trivial task. For this, we apply presentation of 
the object in gray scales. Brighter tones correspond to pixels visiting by 
the representative point with higher probability \cite{23}. In panel (a) this 
technique is used to draw the three-dimensional portrait. Angular 
coordinates $(\varphi ,\theta )$ are plotted in a horizontal plane. The 
third variable plotted along the vertical axis is time, within one full 
period of variation of coefficients in the equations (\ref{eq7}). The picture 
reminds rising and swirling smoke. In the cross-section with a horizontal 
plane $t = 0\,\,(\bmod 6)$ observe a fractal-like formation identical to the 
attractor of the three-dimensional map depicted in angular coordinates in 
Fig.~3~b. One more portrait is shown in the panel (b) on the plane of two 
values of real amplitude $\vert a(t)\vert $ and $\vert a(t - 3)\vert $, 
which relate to instants separated by a half of time period of variation of 
the coefficients in the equations (\ref{eq7}). Here, again one can distinguish the 
fractal-like transverse structure linked with the dynamics of the Plykin 
type. This method of visualization may be appropriate in experiments with 
systems of the class under consideration. 

The Lyapunov exponents $\lambda _i $ of equations (\ref{eq7}) are linked with the 
exponents for the Poincar\'{e} map (see (\ref{eq9}), (\ref{eq11})) by an evident 
relation $\lambda _i = \Lambda _i / T$, where $T = 6$ is the period of 
variation of the coefficients in the equations. The procedure of computation 
of the Lyapunov exponents $\Lambda _i $ is analogous to that used for the 
three-dimensional map. Joint iterations of the map (\ref{eq11}) together with a 
collection of four equations in variations are produced. At each step, 
Gram--Schmidt process is applied to the set of vectors, and normalization of 
them to a fixed constant is performed. Figure 7 present the results 
graphically. The first plot (a) shows four Lyapunov exponents $\Lambda _i $ 
in dependence on parameter $\varepsilon $ at fixed $\mu =1$. In the range 
$\varepsilon < \varepsilon _c \approx 2.03$ one of the exponents is positive 
that means chaos. Among other exponents one is zero (up to numerical 
errors), and two are negative. Note a smooth dependence of the largest 
exponent on the parameter. For larger $\varepsilon $ (strong dissipation 
bringing in during the flow down stages) chaos disappears. The second plot (b) 
shows the Lyapunov exponents versus $\mu $ at fixed $\varepsilon=1$. 
Observe that variation of $\mu $ notably effects only one exponent, which 
corresponds, obviously, to approach of orbits to the invariant sphere. 
Presence of a zero exponent reflects invariance of the equations in respect 
to the overall phase shift. Of course, results of computations agree with 
the data from the previous section: at identical $\varepsilon $ and $\mu $ 
three nonzero exponent are equal, up to numerical errors, to those obtained 
for the three-dimensional map. 

\begin{figure*}[htbp]
\includegraphics[width=5.3in]{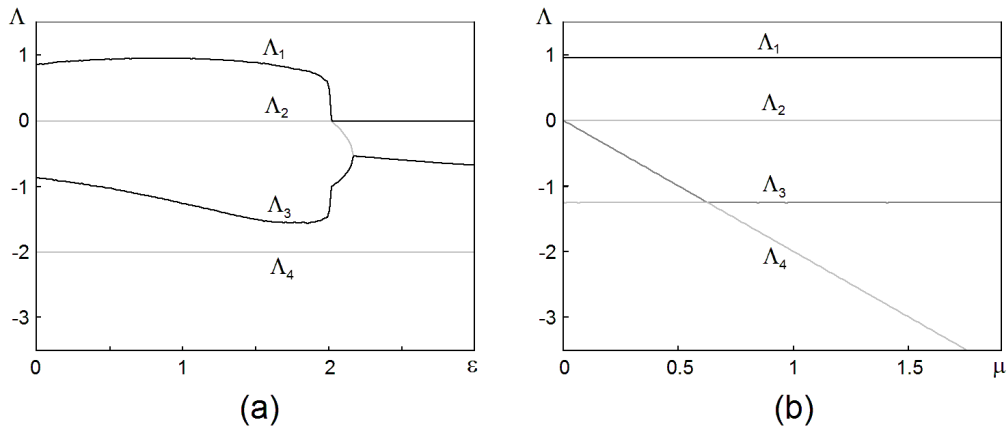}
\label{figur7}
\caption{
Four Lyapunov exponents versus parameter $\varepsilon $ at 
$\mu =1$ (a) and versus parameter $\mu $ at $\varepsilon=1$ (b) for the 
Poincar\'{e} map represented in terms of complex amplitudes. A zero exponent 
occurs due to invariance of the equations in respect to the overall phase 
shift
}
\end{figure*}

Figure 8 shows a plot of spectral density in logarithmic scale versus a 
frequency for a signal generated by one of the oscillators. It relates to 
regime of dynamics on the Plykin -- Newhouse attractor interpreted in terms 
of the three-dimensional map. This spectrum is one more characteristic 
interesting in the context of possible experiments. 
\begin{figure}[htbp]
\includegraphics[width=3.3in]{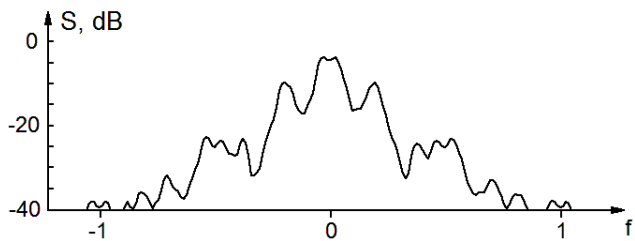}
\label{figur8}
\caption{
The power spectral density for one of the coupled 
oscillators versus the frequency computed by processing a realization 
obtained from numerical integration of the equations (\ref{eq7}) at 
$\varepsilon =1$ 
and $\mu =1$
}
\end{figure}

In computations the standard methodic was used recommended for 
non-parametric statistical estimates of the power spectral density. It is 
based on subdividing the whole realization on a number of parts of equal 
duration. For each part, the signal is multiplied by a smooth function vanishing at 
the ends of the interval (`window'), then Fourier transform is applied, and 
finally, the squared amplitudes of the frequency components are averaged 
over all the parts. A sample of the time series for the complex variable 
$a(t)$ was obtained from numerical solution of equations (\ref{eq7}) by the 
finite-difference method. It corresponds to motion on the attractor and 
contains $6 \cdot 10^5$ data points with time step
$\Delta t = 0.01$. As seen from the 
picture, the spectrum looks continuous that corresponds to chaotic nature of 
the generated signal. The spectrum is almost perfectly symmetric about the 
reference frequency, where the spectral density is maximal. The two main 
side maximums have a level below the central one by about 7 dB, and their 
frequencies approximately correspond to the inverse value of the period of 
variation of the coefficients in equations (\ref{eq7}): $f \approx \pm 1 / 6$. 
Apparently, this periodicity is a reason for the rugged form of the 
spectrum. A plot for the power spectral 
density for the second oscillator looks 
exactly the same because of the symmetry of the system.

\section{Conclusion and discussion}
\label{sec5}

In the present article a system of two coupled non-autonomous nonlinear 
oscillator is introduced manifesting chaotic dynamics, which is in a direct 
relation with the concepts of the hyperbolic theory. With exclusion of the 
overall phase  
the Poincar\'{e} map reduces to a three-dimensional map possessing attractor 
of Plykin type disposed on an invariant sphere. 

As the systems of coupled oscillators occur in many fields in physics and 
technology, it is natural to expect that the suggested model may be 
realizable. Particularly, it may relate to electronic devices, mechanical 
systems, objects of laser physics and nonlinear optics. The systems with 
hyperbolic chaos may be of special interest for applications due to their 
robustness, or structural stability, that means insensitivity of the devices 
to variations of parameters, characteristics of elements, technical 
fluctuations, noise etc. 

Appearance of concrete examples of systems with hyperbolic strange 
attractors makes it reasonable to apply for them the whole arsenal of 
computational methods accumulated in nonlinear dynamics and its 
applications. This is of evident interest both from the point of view of 
complementation of the mathematical concepts with concrete and visible 
context (see e.g. a paper \cite{25}), and for exploiting these concepts in 
applications. In the present work such results of computer studies are 
presented as realizations, attractor portraits, Lyapunov exponents, estimate 
of dimension, power spectral density. 

It is worth mentioning some possible modifications of the model. 
\begin{itemize}
\item
It is easy to suggest a version of the equations, in which the temporal 
dependence of the coefficients will be piecewise smooth. For this, one can 
introduce in the equations 
${\rm {\bf \dot {x}}} = {\rm {\bf f}}({\rm {\bf x}},t)$ 
a smooth common time-dependent factor vanishing at the junctions of the 
stages, integral of which over a stage duration equals 1. An appropriate 
variant is 
${\rm {\bf \dot {x}}} = (2\sin ^2\pi t){\rm {\bf f}}({\rm {\bf x}},t)$. 
The Poincar\'{e} map remains the same, and the nature of the 
attractor is not changed too. (Hunt used a similar trick in his thesis 
\cite{18}.) 
\item
As mentioned, the ``self-oscillatory'' term in the equations proportional to 
$\mu $ may be retained on all stages of the dynamics. 
\item
Working with a version of the model describing by the three-dimensional set 
of equations, it is possible to simplify the form of the nonlinearity 
excluding the operation of extracting the square root, and setting the 
respective term to be $\mu (1 - x^2 - y^2 - z^2)$. This modification does 
not influence dynamics on the attractor belonging to the invariant set 
$x^2 + y^2 + z^2 = 1$.
(In amplitude equations for $a$ and $b$ such 
modification leads rather to complication because of increase of degree of 
the nonlinear terms.)
\item
Taking into account structural stability of the hyperbolic attractor, many 
other modifications can be done, which do not change the nature of the 
attractor, while the variations are not too large. In particular, it is 
possible to introduce a model with smooth analytical variation of the 
coefficients in time in the non-autonomous equations \cite{24}.
\end{itemize} 

Formally, in our complex amplitude equations the attractor should not be 
interpreted as uniformly hyperbolic, because of presence of a neutral 
direction in the state space associated with the overall phase. Instead, it 
has to be related to the class of partially hyperbolic attractors \cite{19,6}. 
Nevertheless, in the form used here invariance of the equations in respect 
to the overall phase is exact; it means that one can accept a rightful 
agreement not to distinguish states distinct only in the phase, and, in this 
sense, treat the dynamics as true hyperbolic. 

However, in systems, for which description in terms of slow complex 
amplitudes will be an approximation, one can expect appearance of 
peculiarities associated with features of the partially hyperbolic 
attractor. If the deflections from the slow-amplitude approximation are 
small, the overall phase will manifest slow random walk, while the dynamics 
of the rest variables will retain its character because of intrinsic 
robustness. 

In dependence on parameters, the suggested model can manifest chaos and 
regular (periodic) dynamics. So, it may serve as an object for principal and 
interesting studies of scenarios of the onset of hyperbolic strange 
attractors in the course of parameter variations (see e.g. \cite{26,27}). 
Insufficient progress in this research direction may be explained 
particularly by the fact that no realistic examples of concrete systems 
undergoing such transitions were known. 

\textit{The work was performed under support of RFBR -- DFG grant No 08-02-91963.}

\begin{flushright}
\textbf{Appendix}
\end{flushright}

For a three-dimensional dissipative map 
${\rm {\bf \bar {x}}} = {\rm {\bf g}}({\rm {\bf x}})$, 
${\rm {\bf x}},\,{\rm {\bf \bar {x}}} \in R^3$ 
let us 
consider the procedure for verification of the cone criterion, bearing in 
mind the situation that one expanding and two contracting directions present 
in the state space, and the attractor is located on the invariant sphere. 
The map is supposed to be reversible: any state vector \textbf{x }has a 
unique image 
${\rm {\bf \bar {x}}} = {\rm {\bf g}}({\rm {\bf x}})$ 
and a unique pre-image 
${\rm {\bf \tilde {x}}} = {\rm {\bf g}^{-1}}({\rm {\bf x}})$.

Let the derivative matrix of the map \textbf{g} at \textbf{x} be 
${\rm {\bf v}} = {\rm {\bf d}}_{\rm {\bf x}} ({\rm {\bf g}})$, 
which acts in the tangent space of vectors 
${\rm {\bf u}} = \left\{ {u_1 ,u_2 ,u_3 } \right\}$. 
Via the auxiliary symmetric matrix 
${\rm {\bf \hat {b}}} = {\rm {\bf v}}^{\rm {\bf T}}{\rm {\bf v}}$, 
where the superscript T means the 
transpose, the norm of the vector 
${\rm {\bf \bar {u}}} = {\rm {\bf vu}}$ 
is expressed as
\begin{equation}
\label{eq16}
\left\| {\rm {\bf \bar {u}}} \right\|^2 = {\rm {\bf u}}^{\rm {\bf T}}{\rm 
{\bf \hat {b}u}}.
\end{equation}
The expanding cone at the point \textbf{x} is a set of vectors 
\begin{equation}
\label{eq17}
S_{\rm {\bf x}} = \left\{ {{\rm {\bf u}}\,\left| {\,{\rm {\bf u}}^{\rm {\bf 
T}}{\rm {\bf \hat {b}u}} \ge \gamma ^2{\rm {\bf u}}^{\rm {\bf T}}{\rm {\bf 
u}}} \right.} \right\}.
\end{equation}
With the same matrix ${\rm {\bf \hat {b}}}$ one can define a pre-image of 
the contracting cone relating to the point ${\rm {\bf \bar {x}}} = {\rm {\bf 
g}}({\rm {\bf x}})$, namely, 
\begin{equation}
\label{eq18}
{C}'_{\rm {\bf x}} = \left\{ {{\rm {\bf u}}\,\left| {\,\gamma ^2{\rm {\bf 
u}}^{\rm {\bf T}}{\rm {\bf \hat {b}u}} \le {\rm {\bf u}}^{\rm {\bf T}}{\rm 
{\bf u}}} \right.} \right\}.
\end{equation}

Now, let us consider an inverse map ${\rm {\bf \tilde {x}}} = {\rm {\bf 
g}}^{ - 1}({\rm {\bf x}})$ and its matrix derivative ${\rm {\bf w}} = {\rm 
{\bf d}}_{\rm {\bf x}} ({\rm {\bf g}}^{ - 1})$. Via the auxiliary symmetric 
matrix ${\rm {\bf \hat {a}}} = {\rm {\bf w}}^{\rm {\bf T}}{\rm {\bf w}}$ we 
represent the norm of the vector ${\rm {\bf \tilde {u}}} = {\rm {\bf wu}}$ 
as
\begin{equation}
\label{eq19}
\left\| {\rm {\bf \tilde {u}}} \right\|^2 = {\rm {\bf u}}^{\rm {\bf T}}{\rm 
{\bf \hat {a}u}}.
\end{equation}
With the matrix ${\rm {\bf \hat {a}}}$ we define the contracting cone at 
\textbf{x} as a set 
\begin{equation}
\label{eq20}
C_{\rm {\bf x}} = \left\{ {{\rm {\bf u}}\,\left| {\,{\rm {\bf u}}^{\rm {\bf 
T}}{\rm {\bf \hat {a}u}} \ge \gamma ^2{\rm {\bf u}}^{\rm {\bf T}}{\rm {\bf 
u}}} \right.} \right\},
\end{equation}
and a cone that is an image of the expanding cone relating to the point 
${\rm {\bf \tilde {x}}} = {\rm {\bf g}}^{ - 1}({\rm {\bf x}})$, namely, 
\begin{equation}
\label{eq21}
{S}'_{\rm {\bf \tilde {x}}} = \left\{ {{\rm {\bf u}}\,\left| {\,\gamma 
^2{\rm {\bf u}}^{\rm {\bf T}}{\rm {\bf \hat {a}u}} \le {\rm {\bf u}}^{\rm 
{\bf T}}{\rm {\bf u}}} \right.} \right\}.
\end{equation}

In computations it is necessary to check, first, existence of nonempty cones 
satisfying the definitions, and, second, validity of the inclusions 
${S}'_{\rm {\bf \tilde {x}}} \subset S_{\rm {\bf x}} $ and 
${C}'_{\rm {\bf \bar {x}}} \subset C_{\rm {\bf x}} $.

As the attractor is placed on an invariant sphere $S^2$, let us assume that 
${\rm {\bf x}} \in S^2$. To define a convenient orthogonal basis $\left\{ 
{{\rm {\bf i}}_1 ,{\rm {\bf i}}_2 ,{\rm {\bf i}}_3 } \right\}$ we take as 
${\rm {\bf i}}_3 $ a unit vector directed along the radius at \textbf{x}, 
and the unit vectors ${\rm {\bf i}}_1 ,\,\,{\rm {\bf i}}_2 $ are taken 
in the tangent plane. To be concrete, we require the matrix element $\hat 
{b}_{12} $ to vanish, and the inequality $\hat {b}_{11} > \hat {b}_{22} $ to 
hold. 

The conditions of required inclusion of the cones are formulated in terms of 
quadratic forms associated with the matrices ${\rm {\bf b}} = {\rm {\bf \hat 
{b}}} - \Gamma ^2{\rm {\bf \hat {e}}}$ and ${\rm {\bf a}} = {\rm {\bf \hat 
{a}}} - \Gamma ^2{\rm {\bf \hat {e}}}$, where ${\rm {\bf \hat {e}}}$ is the 
unite matrix. A constant factor $\Gamma $ is assumed to be equal $\gamma $, 
or 1/$\gamma $, considering the expanding, or contracting cones, 
respectively. Note that the matrices \textbf{a} and \textbf{b} are 
symmetric: $b_{ij} = b_{ji} ,\,\,a_{ij} = a_{ji} $.

Equations 
\begin{equation}
\label{eq22}
b_{11} u_1^2 + b_{22} u_2^2 + b_{33} u_3^2 + 2b_{13} u_1 u_3 + 2b_{23} u_2 
u_3 = 0
\end{equation}
and
\begin{equation}
\label{eq23}
a_{11} u_1^2 + a_{22} u_2^2 + a_{33} u_3^2 + 2a_{12} u_1 u_2 + 2a_{13} u_1 
u_3 + 2a_{23} u_2 u_3 = 0
\end{equation}
determine the borders of the cones. By variable change 
\begin{equation}
\label{eq24}
{u}'_1 = u_1 + b_{11}^{ - 1} b_{13} u_3 ,\,\,\,{u}'_2 = u_2 + b_{22}^{ - 1} 
b_{23} u_3 ,\,\,\,{u}'_3 = u_3 
\end{equation}
the quadratic form in (\ref{eq22}) is reduced to a standard form:
\begin{equation}
\label{eq25}
b_{11} {u'}_1^2 + b_{22} {u'}_2^2 + {b}'_{33} {u'}_3^2 = 0,
\end{equation}
while the equation (\ref{eq23}) becomes
\begin{equation}
\label{eq26}
\begin{array}{c}
a_{11} {u'}_1^2 + a_{22} {u'}_2^2 + {a}'_{33} {u'}_3^2 +\\
+ 2a_{12} {u'}_1{u'}_2 + 2{a}'_{13} {u'}_1 {u'}_3 + 2{a}'_{23} {u'}_2 {u'}_3 = 0.
\end{array}
\end{equation}
Here
\begin{equation}
\label{eq27}
\,\,{b}'_{33} = b_{33} - b_{11}^{ - 1} b_{13}^2 - b_{22}^{ - 1} b_{23}^2 
\end{equation}
and
\begin{equation}
\label{eq28}
\begin{array}{c}
{a}'_{13} = a_{13} - a_{11} b_{11}^{ - 1} b_{13} - a_{12} b_{22}^{ - 1} b_{23}, \\ \\
{a}'_{23} = a_{23} - a_{12} b_{11}^{ - 1} b_{13} - a_{22} b_{22}^{ - 1} b_{23}, \\ \\
{a}'_{33} = a_{33} + a_{11} b_{11}^{ - 2} b_{13}^2 + a_{22} b_{22}^{ - 2} b_{23}^2 
- 2a_{13} b_{11}^{ - 1} b_{13} \\
- 2a_{23} b_{22}^{ - 1} b_{23} + 2a_{12} b_{11}^{ - 1} b_{13} b_{22}^{ - 1} b_{23} . \\ 
 \end{array}
\end{equation}

In the cross-section by a plane $u'_1 = {\rm const}$ 
the equations (\ref{eq25}) 
and (\ref{eq26}) determine some curves of the second order; 
their types and mutual 
location have to be revealed in the course of computations. To have 
situation of inclusion required by the criterion, these curves must be 
ellipses. 

First, in computations we check the inequalities $b_{11} > 0,\,\,\,b_{22} < 
0,\,\,\,{b}'_{33} < 0$. If they are true, the equation (\ref{eq25}) defines an 
ellipse. 

To determine the type of the curve given by (\ref{eq26}), we compute the 
invariants
\begin{equation}
\label{eq29}
I = a_{22} + {a}'_{33} ,\, D = \left| {{\begin{array}{*{20}c}
 {a_{22} } \hfill & {{a}'_{23} } \hfill \\
 {{a}'_{23} } \hfill & {{a}'_{33} } \hfill \\
\end{array} }} \right|,\,\,A = \left| {{\begin{array}{*{20}c}
 {a_{11} } \hfill & {a_{12} } \hfill & {{a}'_{13} } \hfill \\
 {a_{12} } \hfill & {a_{22} } \hfill & {{a}'_{23} } \hfill \\
 {{a}'_{13} } \hfill & {{a}'_{23} } \hfill & {{a}'_{33} } \hfill \\
\end{array} }} \right|
\end{equation}
and check that $D > 0$, and $A / I > 0$. Then, in accordance with the theory 
of conic sections, the equation (\ref{eq26}) also defines an ellipse. 

Let us formulate a convenient and simple sufficient condition of location of 
the second ellipse inside the first one. Renormalizing variables
\begin{equation}
\label{eq30}
\xi = \sqrt {b_{11} } u'_1 ,\,\,\,\eta = \sqrt { - b_{22} } u'_2 
,\,\,\,\zeta = \sqrt { - b_{33} } u'_3 
\end{equation}
and setting $u'_1 = 1 / \sqrt {b_{11} } $, transforms the first ellipse to 
a unit circle
\begin{equation}
\label{eq31}
\eta ^2 + \zeta ^2 = 1.
\end{equation}
Then, the equation for the second ellipse is
\begin{equation}
\label{eq32}
\begin{array}{c}
 - \frac{a_{22} }{b_{22} }\eta ^2 + \frac{2{a}'_{23} }{\sqrt {b_{22} 
{b}'_{33} } }\eta \zeta - \frac{{a}'_{33} }{{b}'_{33} }\zeta ^2  + 
\frac{2a_{12} }{\sqrt { - b_{11} b_{22} } }\eta \\ + \frac{2{a}'_{13} }{\sqrt { 
- b_{11} {b}'_{33} } }\zeta + \frac{a_{11} }{b_{11} } = 0,
\end{array}
\end{equation}          
and its center is located at 
\begin{equation}
\label{eq33}
\begin{array}{l}
\eta _0 = \,\frac{a_{12} {a}'_{33} + {a}'_{13} 
{a}'_{23} }{a_{22} {a}'_{33} - {{a}'_{23}}^2 }\sqrt { - \frac{b_{22} }{b_{11} }}, 
\\ \\
\zeta _0 = \,\frac{{a}'_{13} a_{22} + a_{12} {a}'_{23} }{a_{22} 
{a}'_{33} - {{a}'_{23}}^2 }\sqrt { - \frac{{b}'_{33} }{b_{11} }} .
\end{array}
\end{equation}

In variables 
$\tilde {\eta } = \eta - \eta _0 ,\,\,\tilde {\zeta } = \zeta - \zeta_0 $ 
the equation becomes
\begin{equation}
\label{eq34}
 - \frac{a_{22} }{b_{22} }\tilde {\eta }^2 + \frac{2{a}'_{23} }{\sqrt 
{b_{22} {b}'_{33} } }\tilde {\eta }\tilde {\zeta } - \frac{{a}'_{33} 
}{{b}'_{33} }\tilde {\zeta }^2 = R^2,
\end{equation}
where
\begin{equation}
\label{eq35}
\begin{array}{c}
R^2 =- \frac{a_{11} }{b_{11} }+ \frac{a_{22} }{b_{22} }\eta _0^2 - \frac{2{a}'_{23} }{\sqrt {b_{22} 
{b}'_{33} } }\eta _0 \zeta _0 
+ \frac{{a}'_{33} }{{b}'_{33} }\zeta 
_0^2 \\ \\ - \frac{2a_{12} }{\sqrt { - b_{11} b_{22} } }\eta _0 - \frac{2{a}'_{13} 
}{\sqrt { - b_{11} {b}'_{33} } }\zeta _0 .
\end{array}
\end{equation}
Computing the lesser root of the square equation
\begin{equation}
\label{eq36}
l^2 + (a_{22} / b_{22} + {a}'_{33} / {b}'_{33} )l + (a_{22} {a}'_{33} - 
4{{a}'_{23}}^2 ) / (b_{22} {b}'_{33} ) = 0,
\end{equation}
one finds out the semi-major axis $R / \sqrt {l_{\min } } $. A sufficient 
condition for the ellipse to be located inside the unit disc is inequality
\begin{equation}
\label{eq37}
\sqrt {\eta _0^2 + \zeta _0^2 } \, + R / \sqrt {l_{\min } } < 1.
\end{equation}

If all the named conditions are true at $\Gamma = \gamma $ and at $\Gamma = 1 / 
\gamma $, one can deduce about the correct inclusion for the 
expanding and contracting cones at the analyzed point \textbf{x}. 

Indeed, in variables $\eta $, $\zeta $ the cross-section of the expanding 
cone $S_{\rm {\bf x}} $ is the closed unit disc, and cross-section of the 
cone ${S}'_{\rm {\bf \tilde {x}}} $ is represented by the closure of 
interior of the small ellipse obtained at $\Gamma = \gamma $, so, the 
required inclusion ${S}'_{\rm {\bf \tilde {x}}} \subset S_{\rm {\bf x}} $ 
is valid (Fig.~9~a). On the other hand, cross-section of the contracting 
cone $C_{\rm {\bf x}} $ is a closure of exterior of the small ellipse 
obtained at $\Gamma = 1 / \gamma $. Cross-section of the cone ${C}'_{\rm 
{\bf \bar {x}}} $ corresponds to a closure of exterior of the unit circle. 
Hence, the inclusion ${C}'_{\rm {\bf \bar {x}}} \subset C_{\rm {\bf x}} $ 
is valid (Fig.~9~b).

\begin{figure}[htbp]
\includegraphics[width=3in]{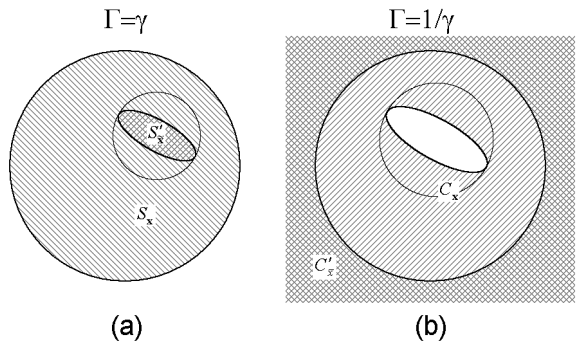}
\label{figur9}
\caption{
Cross-sections of the cones for the case of 
three-dimensional map with one expanding and two contracting directions. A 
picture of proper inclusion is shown (a) for expanding cones 
${S}'_{\rm {\bf \tilde {x}}} \subset S_{\rm {\bf x}} $ 
and (b) for contracting cones 
${C}'_{\rm {\bf \bar {x}}} \subset C_{\rm {\bf x}} $ 
A circle circumscribed around 
the ellipse is located inside the unit disc if the condition (19) is valid
}
\end{figure}

Computations in the present work were organized as verification of the cone 
criterion for a set of points on the attractor obtained from multiple 
iterations of the map ${\rm {\bf g}}({\rm {\bf x}})$. 
The inclusions were checked with
a help of the inequality (\ref{eq37}).

Because of smoothness of the map under study, the objects considered in the context of 
the cone criterion (matrices, quadratic forms and their invariants) depend 
on the state variables in smooth manner, as they are determined by dynamics 
on finite time intervals. As follows, validity of the conditions at some 
point \textbf{x} with distant from 1 constant $\gamma $ implies that the 
cone criterion holds as well in a neighborhood of \textbf{x} (as wider, as 
larger the value $\vert \gamma - 1\vert $ is). A positive result of the test 
for a representative set of points implies validity of the cone criterion on 
the whole attractor, if it is completely covered by the union of the 
mentioned neighborhoods. Practically, such situation is achieved by increase 
of the number of iterations and, respectively, a number of tested points on 
the attractor. 

It is convenient not to fix in advance the constant $\gamma $, but to 
arrange the computations as follows. First, at each point \textbf{x} we 
compute the matrices ${\rm {\bf \hat {a}}}$ and ${\rm {\bf \hat {b}}}$ and 
check all the formulated conditions for $\gamma = 1$. If they hold, the 
program determines an allowable interval of $\gamma $. For this, the program 
simply enumerates the $\gamma $ values with a small step in a sufficiently 
wide range. Attractor is recognized as hyperbolic if a gap of a finite width 
separates the obtained sets of top and bottom edges of the intervals from 
the axis $\gamma = 1$ on the plot of $\gamma $ versus some dynamical 
variable characterizing location of the analyzed point \textbf{x}.

\end{document}